\documentclass[
  ,aps
  ,twocolumn
  ,superscriptaddress
  ,showkeys
  ,showpacs
  ,10pt
  ]
  {revtex4}

\usepackage{amsmath}
\usepackage{graphicx}
\usepackage{dcolumn}
\usepackage{bm}
\usepackage{enumerate}
\usepackage{enumitem}
\usepackage{ctable}
\usepackage{booktabs}
\usepackage{natbib}
\usepackage{siunitx}
\usepackage{pifont}

\begin{document}

\title{Developing the Next Generation of Physics Assessments}


\author{James T. \surname{Laverty}}
\affiliation{CREATE for STEM Institute, Michigan State University, East Lansing, MI 48824, USA}

\author{Melanie M. \surname{Cooper}}
\affiliation{Department of Chemistry, Michigan State University, East Lansing, MI 48824, USA}
\affiliation{CREATE for STEM Institute, Michigan State University, East Lansing, MI 48824, USA}

\author{Marcos D. \surname{Caballero}}
\affiliation {Department of Physics and Astronomy, Michigan State University, East Lansing, MI 48824, USA}
\affiliation {CREATE for STEM Institute, Michigan State University, East Lansing, MI 48824, USA}

\begin{abstract}

Science education at all levels is currently undergoing dramatic changes to its curricula and developing assessments for these new curricula is paramount.  We have used the basis of many of these new changes (scientific practices, crosscutting concepts, and core ideas) to develop sets of criteria that can be used to guide assessment development for this new curriculum.  We present a case study that uses the criteria we have developed to revise a traditional physics assessment item into an assessment item that is more aligned with the goals of current transformation efforts.  Assessment items developed using this criteria can be used to assess student learning of both the concepts and process of science.


\end{abstract}


\maketitle


\section{Introduction}


The science education curriculum is currently undergoing a significant overhaul in the ways students are taught and learning is assessed. In 2012, the National Research Council released {\it A Framework for K-12 Science Education: Practices, Crosscutting Concepts, and Core Ideas} (the {\it Framework}), which described, in detail, three dimensions of science and engineering learning on which to base a curriculum\cite{Standards:2012ur}. 
The three dimensions are briefly described here, followed by examples from the document:

\begin{description}[noitemsep]
\item[Scientific Practices] These eight practices can be thought of as the disaggregated components of the process of science. These practices engage students in science by putting their knowledge to use to model, predict, and explain phenomena. {\it Examples: Planning and Carrying Out Investigations; Constructing Explanations}
\item[Crosscutting Concepts] These seven concepts bridge the boundaries between the disciplines of the physical, biological, and geological sciences. These ``ways of thinking'' are used by each of the disciplines and can be leveraged to help students make connections across them. {\it Examples: Patterns; Cause and Effect: Mechanism and Explanation}
\item[Disciplinary Core Ideas] These are the foundationally important concepts that are fundamental to members of a scientific discipline.  In order to qualify as a disciplinary core idea, the concept must be essential to the study of the discipline, be required to explain a wide range of phenomena, and provide a way to generate new ideas and predictions. {\it Examples: Matter and its Interactions; Energy}
\end{description}

The {\it Framework} emphasizes that it is vital that all three dimensions are blended into nearly every aspect of students' learning opportunities (which we call ``three-dimensional learning'').


\subsection{Bringing Three-Dimensional Learning to College Classrooms}

While the {\it Framework} was written for the K-12 community, its conclusions (which were informed by both science education and discipline-based education research) can and should be applied to introductory college science courses as well\cite{Cooper15ScienceForum}. After all, college instructors certainly would like to help guide their students to, for example, be able to construct explanations (scientific practice) for systems (crosscutting concept) that utilize the concept of energy (disciplinary core idea).


With the adoption of the Next Generation Science Standards (which are based on the {\it Framework}) by more and more states, the K-12 education system is currently heading in the direction of implementing three-dimensional learning at the national scale\cite{NGSS}.  In our view, universities and colleges can and should react to this for two reasons.  First, as time passes, more students entering college will have experienced three-dimensional learning.  If universities and colleges aim to capitalize on this new form of science education, we must start adapting and adopting now. Second, with the significant effort needed to incorporate the new standards, K-12 teachers will need to be trained not only in the knowledge of science, but also its process.  These teachers are often required to take introductory college science courses, which means we have an opportunity (duty?) to help develop the next generation of science teachers.

\subsection{Assessing Three-Dimensional Learning}


At the college level, assessments have helped drive transformation efforts and measure their impact in the past.  In fact, much of the early history of discipline-based education research focused on the development of assessments and adjusting curricula to help students improve on those assessments\cite{Hestenes:1992ws,Maloney:2001cx}. These `concept inventories' are often suggested and used as catalysts or assessments for efforts to transform teaching and learning\cite{DBERreport}.  Thus, being able to assess three-dimensional learning should be a high priority for modifying instruction. After all, if we do not assess what is important, then what is assessed becomes important.  Additionally, as assessments are a part of students' learning opportunities, they often drive the `hidden curriculum', which has a significant effect on student learning\cite{1998AmJPh..66..212R}.

The remainder of this paper will introduce work that endeavours to help instructors (and other assessment authors) develop assessment items (e.g., exam, homework, and clicker questions) that align with three-dimensional learning.  Given the format for this paper, we will present only a single case study, specifically for a college level physics course. To do this, first we introduce criteria from a newly developed instrument, then use these criteria to analyze a typical introductory physics problem, and finally modify that problem to better align with the three dimensions.


\section{Developing Three-Dimensional Assessments}

\begin{table*}[t]
\caption{An analysis of the assessments in this article using the criteria we have developed for the 3D-LAP instrument. An assessment item must meet all of the criteria for a dimension in order for it to be considered `Aligned' with that dimension.}
\begin{tabular}{ p{5.5cm} | p{5.5cm} | p{5.5cm} }

\textbf{3D-LAP criteria for aligning with three-dimensional learning} &\textbf{Elements of Traditional Item (Figure \ref{fig:traditional}) that meet the criteria} 		&\textbf{Elements of Revised Item (Figure \ref{fig:tetherball}) that meet the criteria} \\
\hline
\underline{Developing and using models:} 	&\underline{Unaligned}	&\underline{Aligned} \\

1. Question gives an event, observation, or phenomenon for the student to explain or make a prediction about. 	&\ding{55} 1. Question describes a system that is already idealized.	&\ding{51} 1. Question stem describes ``the faster the ball moves, the larger the angle the rope makes with the pole.'' \\

2. Question gives student a representation or asks student to select a representation. 	&\ding{51} 2. Question includes a visual representation of the situation.	&\ding{51} 2. Question 1 asks ``Which of the following free body diagrams''  \\

3. Question asks student to select an explanation for or prediction about the event, observation, or phenomenon using the representation. 	&\ding{51} 3. Question asks ``what is the net force acting on the mass?''	&\ding{51} 3. Answer 3a includes ``As the speed increases, the angle must increase.'' \\

4. Question asks student to select the reasoning that links the representation to their explanation or prediction. 	&\ding{55} 4. Question does not ask for any evidence of reasoning.	&\ding{51} 4. Answer 1b (free body diagram), answer 2a (equation), and answer 3a includes ``in order for the net force to maintain the ball's circular trajectory.'' \\
\hline

\underline{System and system models:} 	&\underline{Unaligned}	&\underline{Aligned}\\

1. The components of the system. 	&\ding{51} 1. Question asks ``acting on the mass?''	&\ding{51} 1. Question 1 asks ``Which of the following free body diagrams would apply to the ball'' \\

2. The interactions between those components. 	&\ding{51} 2. {\it Does not apply, since there is only one object in the system.}	&\ding{51} 2. {\it Does not apply, since there is only one object in the system.} \\

3. The interactions of those components with the surroundings. 	&\ding{55} 3. Question and answer do not include evidence of the interactions.	&\ding{51} 3. Answer 1b represents the interactions with the environment. \\

\end{tabular}
\label{table:criteria}
\end{table*}

\subsection{Assessment Item Criteria}

As part of a larger transformation project at Michigan State University, we have been developing the Three-Dimensional Learning Assessment Protocol (3D-LAP), a soon-to-be released research instrument that can be used to both design and characterize assessment items.  The principle components of the 3D-LAP are sets of criteria that have been developed for each scientific practice, crosscutting concept, and core idea.  These criteria can be used by assessment authors to help modify or generate items that align with the three dimensions of the {\it Framework}.  Items that meet these criteria align with the scientific practices, crosscutting concepts, and core ideas as well as give instructors more explicit feedback about what students are capable of doing.

The 3D-LAP uses the list of scientific practices and the list of crosscutting concepts from the {\it Framework}.
Because there is no national document that has a list of core ideas for college level physics, we are left to invent our own.  At Michigan State University, faculty in the Physics and Astronomy department have discussed and generated the set of core ideas for our courses\cite{LavertyPERC14}.

There is a fundamental difference between the three dimensions that results in different formats for the criteria.  Because the scientific practices are about the process of science, they require that certain actions be performed (`verbs').  As a result, the criteria for the scientific practices all focus on the activities the assessment item asks the student to engage in.  Unlike the scientific practices, crosscutting concepts and core ideas are constructs that represent scientific knowledge (`nouns').  As such, the criteria for them include subconstructs that must either be indicated in the question, or asked of the student (in our example this means it must be part of the question stem or the correct answer choice.)  Sample criteria from the 3D-LAP can be found in the first column of Table \ref{table:criteria}.


\subsection{A Traditional Assessment Item}

\begin{figure}[t]
\fbox{
\includegraphics[width=.9\columnwidth]{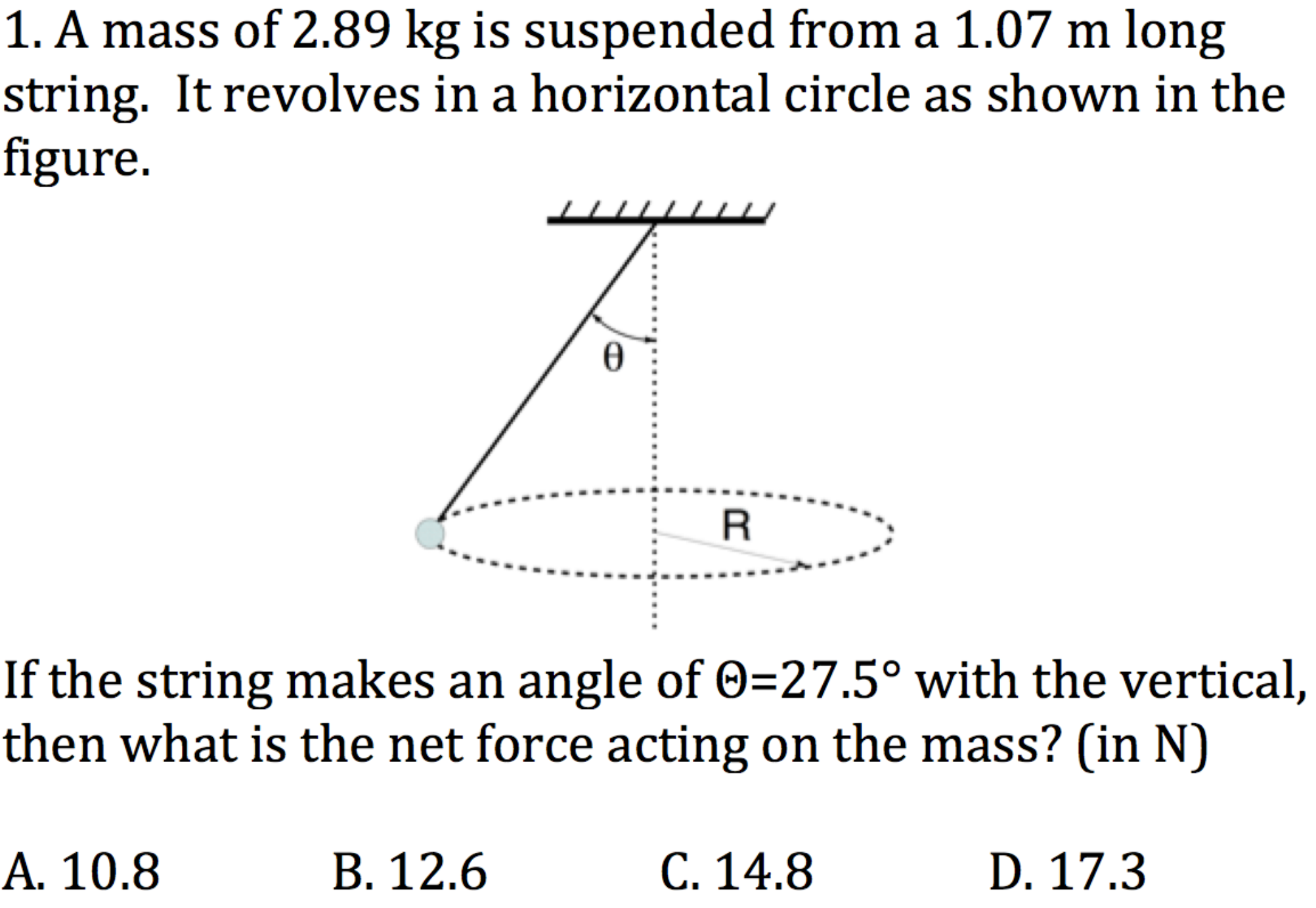}
}
\caption{A traditional introductory physics assessment item that aligns with one of the three dimensions in the {\it Framework}.}
\label{fig:traditional}
\end{figure}

Figure 1 shows a typical introductory physics question for which an instructor might hope students would build a free body diagram, use Newton's second law to generate a mathematical representation, and solve those equations for a particular value.  Unfortunately, this item does not, in fact, give explicit evidence that a student is doing any of those things, as choosing the correct or incorrect solution (in this case) does not help us understand how a student arrived at that particular answer.

In order to accomplish this and, in the process, align with each of the dimensions of the {\it Framework}, we must first decide which scientific practice, crosscutting concept, and core idea this question already aligns with most closely (or, alternatively, which ones we would like it to align with).  For this example, we have chosen:

\begin{description}[noitemsep]
\item[Scientific Practice] Developing and using models
\item[Crosscutting Concept] Systems and system models
\item[Disciplinary Core Idea] Interactions can cause changes in motion
\end{description}

The central column of Table \ref{table:criteria} contains our analysis of the question in Figure \ref{fig:traditional}, including evidence for why it does or does not meet each particular piece of the criteria.

In this particular case, the question in Figure \ref{fig:traditional} is most closely connected to our (locally determined) core idea of {\it Interactions can cause changes in motion}\cite{LavertyPERC14}.  While one might argue that this is not the proper core idea, or that they would use a different one, we strongly suspect that every such list would include the idea of forces, interactions, and their effects on the motions of objects.  For this reason, we argue that this traditional item already contains a core idea and do not include the criteria or analysis in Table 1 due to space considerations.

\subsection{Revising an Assessment Item}

\begin{figure}[t]
\fbox{
\includegraphics[width=.9\columnwidth]{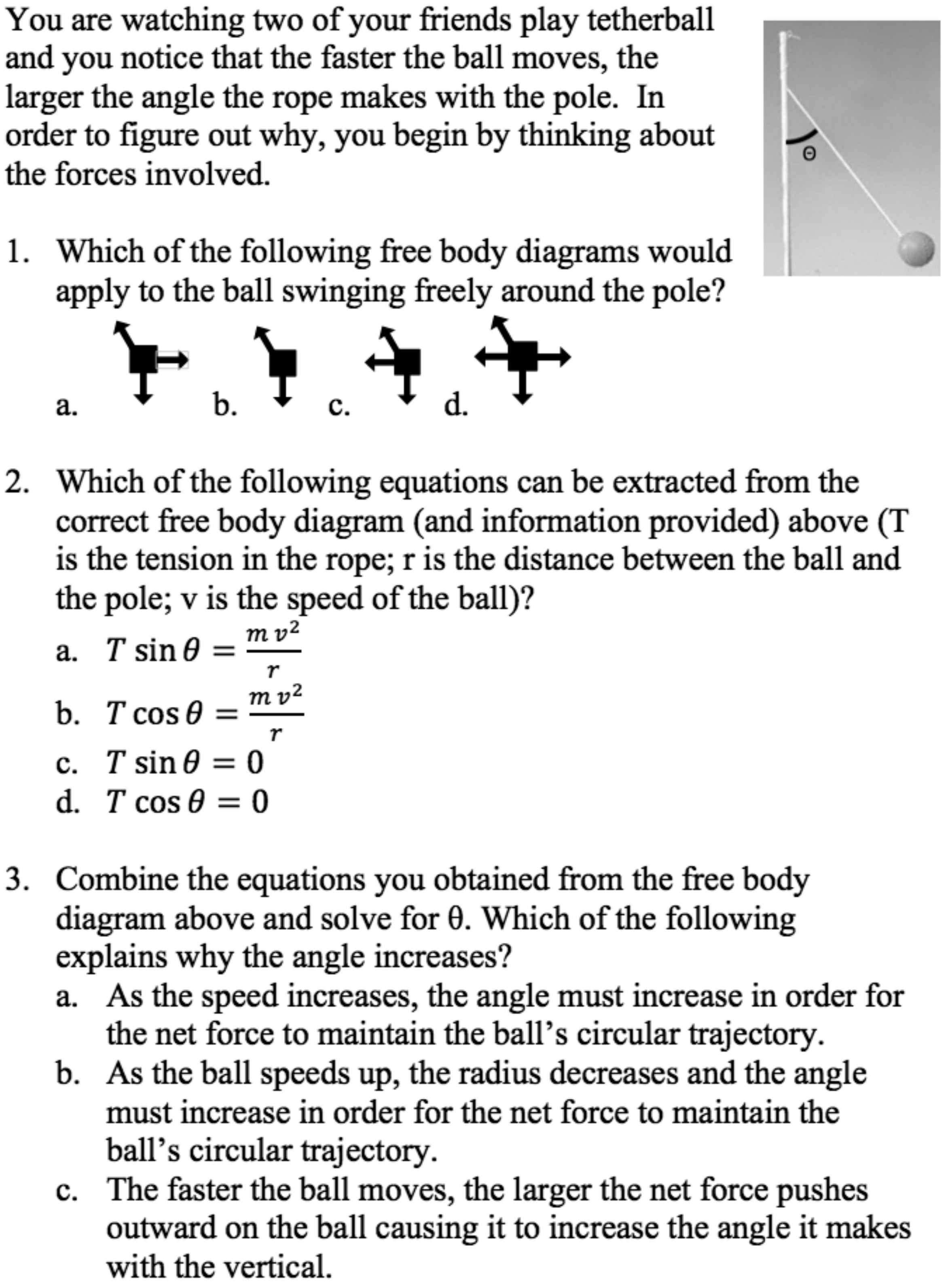}
}
\caption{The revised cluster of assessment items modified to align with the three dimensions using the criteria from the 3D-LAP.}
\label{fig:tetherball}
\end{figure}

In order to obtain more explicit evidence of student reasoning, the revised item was developed into a cluster of three related questions. Figure \ref{fig:tetherball} contains the revised version of the assessment item, which we claim aligns with the three dimensions.  The analysis of how it aligns with the chosen scientific practice and crosscutting concept can be found in the right most column of Table \ref{table:criteria}.


For the scientific practice, the original question is not meeting two of the criteria.  First, the situation described is already idealized (moving in a perfect circle with no friction).  One part of the modification must be to place the question in a real world setting (which might then be modeled as movement in a circle with no friction).  The second missed criterion requires evidence about the reasoning the student is using to answer the question.  Because this is a selected response setting, multiple options for the reasoning students might give must be included as part of the set of answers.  

As for the crosscutting concept, the original question does not include any information about the interactions between the system and its surroundings.  That information must be provided in some way or another. For this particular revision, we have chosen to have the student select the appropriate free-body diagram, which contains this information. 

Ultimately, we claim (based on the analysis found in Table \ref{table:criteria}) that the revised cluster of items in Figure \ref{fig:tetherball} is more capable of assessing a student's ability to engage with a scientific practice, crosscutting concept, and core idea than the traditional item in Figure \ref{fig:traditional}.

\section{Discussion}

Comparing the traditional and revised assessment item, one might argue that we have altered the intent of the question because we no longer ask the student to solve numerically for a particular value.  However, it is important to note that while such a question is not necessary to meet the criteria, it could still be asked.  In fact, one could imagine adding the traditional question as question number 4 in the revised cluster and the cluster would still meet all of the criteria to qualify as three-dimensional.  This points to an interesting question about assessing three-dimensional learning: What about items that we want to assess students with that do not include all three dimensions? We do not claim that all assessment items students engage with need to align with all three dimensions, but clearly some of them need to if we are going to claim that we are assessing three-dimensional learning.

Assessing how a student authentically engages in scientific practices requires that they generate their own ideas in response to a situation where there are no clear answers.  Selected response items generally do not work well for probing student engagement with scientific practices because they often require that one answer choice is clearly correct over the other options, something uncommon in science.  For this reason, we maintain that only constructed response items can engage students in scientific practices.

However, because many introductory science courses are large, constructed response assessment items are often prohibitively expensive and/or time-consuming to grade.  Selected response items, on the other hand, require fewer resources to grade, increasing the potential for both faculty productivity and quick turnaround of feedback.  For these reasons, assessment items in selected response format are often necessary for these courses.

For the purposes of this paper, we felt it was important to show the modification of a selected response assessment item without changing it to a constructed response format, as this is the more difficult challenge.  Given the argument above, we do not claim that meeting the criteria in this case engages students in a scientific practice, but that it is as close to aligning with the practice as it can be while remaining a selected response item.

While reviewing assessment items during the development of the criteria for the 3D-LAP, one common theme we discovered was that many traditional assessment items (like the one in Figure \ref{fig:traditional}) meet at least some of the criteria for scientific practices.  However, many of the scientific practices criteria require evidence of some kind of reasoning (similar to criterion 4 under {\it Developing and using models} in Table \ref{table:criteria}) and this particular criterion is almost never met by traditional assessment items.

\section{Conclusion}

Being able to engage in both the process of science (scientific practices) and the knowledge of science (crosscutting concepts, core ideas) should be the goal of science education.  Research is beginning to show that the framework of three-dimensional learning is a productive way to go about designing curricula and assessments to help students accomplish these goals.  For this reason and those described earlier, it is necessary to bring three-dimensional learning to our college level physics courses.

If we are going to include it, we need a way to assess it and the criteria we have developed can help to develop the kinds of assessments necessary to promote student learning.  In the near future, we will be publishing the 3D-LAP for others to use, critique, and hopefully improve upon so that we can collectively move to the next generation of physics assessments.

\section{Acknowledgements}

The authors thank the Association of American Universities' STEM Education Initiative for their support, the rest of the DBER community at MSU (especially those involved in the development of the criteria), and PERL@MSU for their feedback on early drafts.

\bibliographystyle{apsrev}
\bibliography{thehitlist}

\end{document}